\documentclass[aps,prB,amsmath,amssymb,twocolumn,groupedaddress,notitlepage,showpacs,floatfix,superscriptaddress]{revtex4-1}

\usepackage{graphicx}
\usepackage{bm,color}
\usepackage{physics}
\usepackage{amsmath}
\usepackage{bm,color}
\usepackage{multirow}

\usepackage{graphicx}
\usepackage{bm,color}

\usepackage{amsmath}
\usepackage{bm,color}
\usepackage{multirow}
\usepackage{ulem}
\usepackage{xcolor}

\newcommand{\be}{\begin{eqnarray}}
\newcommand{\ee}{\end{eqnarray}}

\newcommand{\ver}{\vec{r}}
\newcommand{\verz}{\vec{0}}
\newcommand{\hi}{\hat{i}}

\newcommand{\hx}{\hat{x}}
\newcommand{\hy}{\hat{y}}
\newcommand{\CB}{{\cal B}}
\newcommand{\CS}{{\cal S}}
\newcommand{\IT}{e^{i\theta}}

\begin{document}

\title{
Flat-band full localization and symmetry-protected topological phase on bilayer lattice systems}
\date{\today}
\author{Ikuo Ichinose}
\affiliation{Department of Applied Physics, Nagoya Institute of Technology, Nagoya, 466-8555, Japan}
\author{Takahiro Orito}
\affiliation{Quantum Matter program, Graduate School of Advanced Science and Engineering, Hiroshima University, Higashi-Hiroshima 739-8530, Japan}
\author{Yoshihito Kuno}
\affiliation{Department of Physics, Graduate School of Science, Tsukuba University, Tsukuba, Ibaraki 305-8571, Japan}

\begin{abstract}
In this work, we present bilayer flat-band Hamiltonians, in which all bulk states are localized and 
specified by extensive local integrals of motion (LIOMs).
The present systems are bilayer extension of Creutz ladder,
which is studied previously.
In order to construct models, we employ building blocks, cube operators, which are linear combinations of fermions
defined in each cube of the bilayer lattice.
There are eight cubic operators, and the Hamiltonians are composed of the number operators of them, the LIOMs.
A suitable arrangement of locations of the cube operators is needed to have exact projective Hamiltonians. 
The projective Hamiltonians belong to a topological classification class, 
BDI class. 
With the open boundary condition, the constructed Hamiltonians have gapless edge modes, 
which commute with each other as well as the Hamiltonian.
This result comes from a symmetry analogous to the one-dimensional chiral symmetry of the BDI class. 
These results indicate that the projective Hamiltonians describe a kind of 
symmetry protected topological phase matter.
Careful investigation of topological indexes, such as Berry phase, string operator, is given.
We also show that by using the gapless edge modes, a generalized Sachdev-Ye-Kitaev (SYK) model is constructed.
\end{abstract}


\maketitle
\section{Introduction}
Flat-band systems are one of the most attractive topics in condensed matter community. 
Such systems exhibit exotic localization phenomena without disorder, currently called disorder-free 
localization \cite{Smith1,Smith2,WSL2019,McClarty,Scherg}.
Especially, the system, which is totally composed of flat-bands and called complete flat-band system, generally 
possesses extensive number of local conserved quantities called local integrals of motion 
(LIOMs) \cite{Nandkishore,Abanin,Imbrie,Serbyn}. 
Without interactions, complete flat-band systems are, therefore, integrable and their dynamics exhibits
non-thermalized behaviors \cite{Mukherjee,Vidal0,Naud}.
The idea of LIOMs was firstly introduced in the study of many-body localization (MBL) \cite{Nandkishore,Abanin,Imbrie,Serbyn}. 
There, emergent LIOMs induce localization, non-thermalized dynamics with slow increase of 
entanglement entropy \cite{Bardarson}.
On the other hand in the complete flat-band systems, the origin of the LIOMs is due to the presence of
the compact localized states (CLS) \cite{Flach,Mizoguchi2019,KMH2020,KMH2020_2}, 
hence the LIOMs are explicitly given in terms of the number operator of the CLS.
The origin of LIOMs in these systems is essentially different from that of the emergent LIOMs in the MBL systems, but both of them play an important role
concerning to localization. 

From another point of view, flat-band systems with nontrivial topological bands have attracted many interests. 
On flat-bands, the kinetic terms are negligible and interactions play a dominant role in determining
the ground state.
Then, such a system possibly exhibits exotic topological phases.  
Fractional Chern insulator is expected to be realized in nearly flat-band systems \cite{Regnault,Bergholtz}, 
and for complete flat-band systems, fractional topological phenomena have been reported \cite{Guo,Budich,Barbarino_2019}. 
Also, flat-band system is closely related to frustrated systems, in which
huge degeneracies exist in the vicinity of the ground state, and as a result, 
some kinds of topological phases possibly emerge 
there \cite{Tomczak,Richter,Derzhko,Wildeboer}.
Recently, some interesting works on flat-band systems with nontrivial topological bands have been reported~\cite{Jiang1,Jiang2,Jiang3}.

As a typical example of such a flat-band system, Creutz ladder \cite{Creutz1999}
with a fine-tuning is an interesting system, where the two complete flat-bands and
the two types of CLS appear. 
Due to the presence of the extensive number of the CLS, the model exhibits explicitly disorder free localization phenomena, called Aharanov-Bohm caging \cite{Mukherjee,Vidal0,Naud}.
It is known that localization tendency survives even in the presence of 
interactions \cite{KOI2020,OKI2020,Roy,Danieli_1,OKI2021}. 
Furthermore, Creutz ladder shows some topological phases with quantized Berry phase and zero energy edge states \cite{Creutz1999,Bermudez,Junemann,Zurita,Sun,YK2020}, 
and interestingly fractional topological phenomena \cite{Barbarino_2019}. 
However, except for Creutz ladder, complete flat-band models with
nontrivial topological properties have not been studied in great detail so far. 
Hence, exploring such a model that goes beyond the above example remains an open issue.

In the present paper, by extending the character of the CLS in Creutz ladder, 
we propose novel types of flat-band systems on bilayer lattice, 
which can be set in both one and two-dimensional (2D) lattice geometries.
There, the extensive LIOMs are explicitly obtained
and all states are localized in the periodic boundary condition.
As a topological aspect, the system Hamiltonians have symmetries of the BDI class 
in {\it ten-fold way} \cite{Altland,Ludwig,Chiu,Ryu}. 
With BDI symmetry, the system Hamiltonian 
set on quasi-1D lattice explicitly exhibits symmetry protected topological (SPT)
phase \cite{Pollmann2010,Chen,Pollmann2012}.  
Also, we find that, in both 1D and 2D systems with open boundary conditions,
there emerge gapless edge modes as a result of chiral symmetry. 
The gapless edge modes can be analytically given due to the presence of the CLS.
Therefore, the presence of the edge modes implies that the present systems can 
be regarded 
as a 2D SPT system whose bulk states are full localized.   
Such kind of models in 1D is studied, e.g., in Ref.~\cite{Bahri}.

This paper is organized as follows.
In Sec.~II, we prepare building blocks called cube operators that are used for the construction of the model Hamiltonians.
There are eight kinds of cube operators, which are linear combination of fermions located on eight sites of a cube.
The cube operators located on the same cube commute with each other, but certain pairs of them
located on next-nearest-neighbor (NNN) cubes do not. 
The cube operators transform with each other by 
chiral-symmetry transformation, and we require the invariance
of the Hamiltonian under chiral-symmetry transformation.  
In Sec.~III, we construct models and their Hamiltonian by using the cube operators.
Certain specific location of eight kinds of the cube operators is required to obtain exact projective
Hamiltonian.
We present two kinds of such Hamiltonian, one of which is defined on a bilayer lattice and the other
on a square prism lattice.
Interactions between fermions can be introduced, which are expressed by the LIOMs and satisfy chiral symmetry.
Section IV is devoted for study on the edge modes in the above two models.
As a result of chiral symmetry of the bulk Hamiltonian, the edge modes are invariant under the transformation
corresponding to chiral symmetry.
Furthermore, effective Hamiltonian of the edge modes is derived, which is an extension of the SYK model~\cite{SachdevM,YeM,KitaevM}.  
Section V is devoted for discussion on topological indexes, which characterize non-trivial topological
properties of the emergent eigenstates.
Numerical study of the square prism model is given to examine the stability of the topological states.
Section VI is devoted for conclusion and discussion.
We explain that flat-band localization by the CLS plays an important role for topological properties of the systems.
 
\section{Construction of bilayer models}

In the previous works~\cite{KOI2020,OKI2020,OKI2021}, we studied fermion systems on Creutz ladder,
and obtained interesting results concerning to flat-band localization and topological phase.
In this section, we construct fermion systems on the bilayer lattice that exhibits the full-localization
and topological properties.
These systems have projective Hamiltonian with time-reversal (${\cal T}$), particle-hole (${\cal C}$) and chiral symmetries (${\cal S}={\cal T}{\cal C}$).
As a result, they have gapless edge modes under the open boundary condition (OBC), whereas the bulk states
are full localized and have flat-band dispersion.
To construct models, we prepare eight building blocks with the cubic shape, each edge of which corresponds
to a linear combination of fermions at two sites of the edge,  which is an extension of the CLS in Creutz ladder~\cite{Creutz1999,Bermudez,Junemann,Zurita}.
Using these building blocks called the cube operators, we can construct bilayer models with various shapes,
including a torus, a thin cylinder, and a square prism.
In the context of the study of MBL, each cube operator is also regarded as a $\ell$-bit (equivalent to the CLS)
\cite{Flach}, and the target Hamiltonians are obtained by using
LIOMs, which are nothing but the number operator of the $\ell$-bits (CLS). 



\begin{figure}[t]
\begin{center} 
\includegraphics[width=7cm]{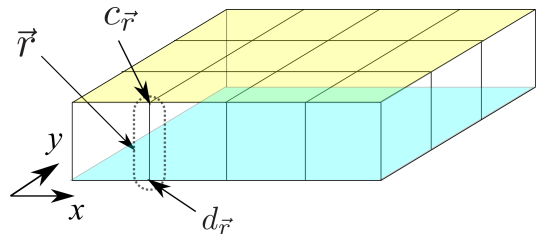} 
\end{center} 
\caption{
Schematic picture of fermion operators on bilayer lattice.
$c_{\ver}$'s reside on upper square lattice, and $d_{\ver}$'s on lower lattice,
where $\ver=(x,y)$.
}
\label{Fig1}
\end{figure}

\subsection{Eight cube operators}

Let us consider a cube, which is a unit cell of the bilayer lattice, i.e., each vertex of the cube is located 
at a lattice site, see Fig.~\ref{Fig1}.
We introduce ($x$-$y$-$z$) axes, and fermion creation (annihilation) operators $c^\dagger_{\ver}$ 
$(c_{\ver})$ and
$d^\dagger_{\vec{r}}$ $(d_{\ver})$, where $\vec{r}$ denotes lattice sites, $\vec{r}=(x,y,z)$, and $z=1 \; (2)$
for $c_{\ver} \; (d_{\ver})$.
Therefore, the fermion $c_{\vec{r}}$ and $d_{\vec{r}}$ are located in the upper and lower layer, respectively,
and then we shall use $\ver=(x,y)$ hereafter.


\begin{figure*}[t]
\begin{center} 
\includegraphics[width=17cm]{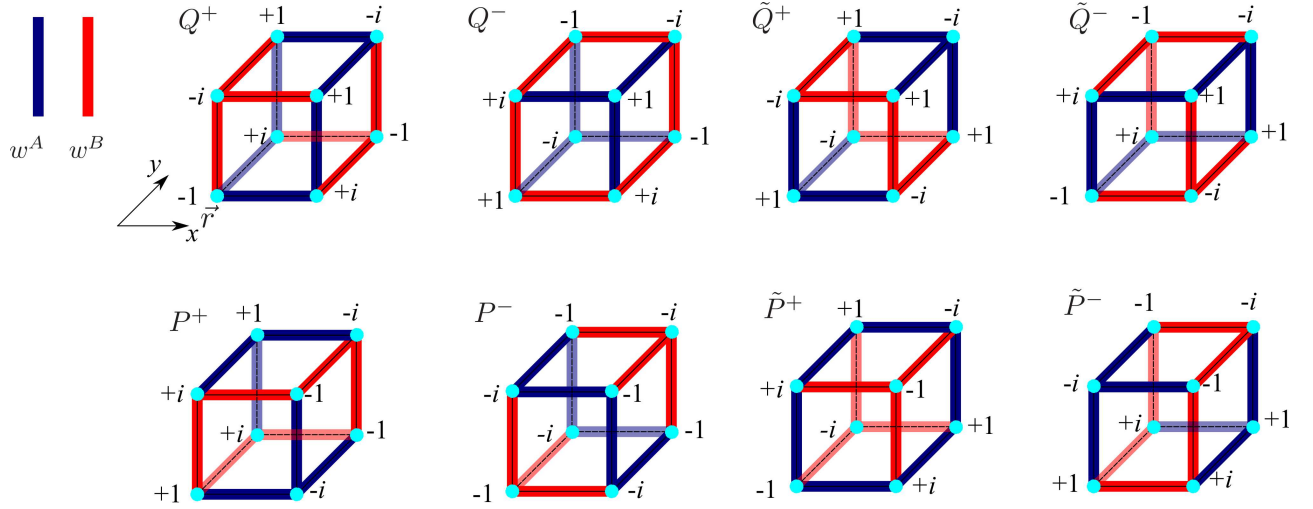} 
\end{center} 
\caption{
Schematic picture of eight cube operators on bilayer lattice, $Q^+_{\ver} \sim \tilde{P}^-_{\ver}$.
Eight cube operators are linear combinations of 
$\{\omega^A_{\ver,\hat{i}},\tilde{\omega}^A_{\ver, \hat{i}}, \bar{\omega}^A_{\ver, \hat{i}}\}$ [blue]
and $\{\omega^B_{\ver,\hat{i}},\tilde{\omega}^B_{\ver, \hat{i}}, \bar{\omega}^B_{\ver, \hat{i}}\}$ [red].
The numbers near vertices of the cube indicates the value of the coefficient of the cube operators.
}
\label{Fig2}
\end{figure*}

The eight cube operators are constructed by $c_{\vec{r}}$ and $d_{\vec{r}}$.
To this end, as elementary building blocks, the following notations are useful;
\be
&&\omega^A_{\ver, \hat{i}}=c_{\ver+\hi}+ic_{\ver}, \;
\omega^B_{\ver, \hat{i}}=c_{\ver+\hi}-ic_{\ver},  \;\; \mbox{(upper layer)},  \nonumber \\
&&\tilde{\omega}^A_{\ver, \hat{i}}=d_{\ver+\hi}+id_{\ver}, \;
\tilde{\omega}^B_{\ver, \hat{i}}=d_{\ver+\hi}-id_{\ver},  \;\; \mbox{(lower layer)},\nonumber \\
&&\bar{\omega}^A_{\ver, \hat{z}}=c_{\ver}+id_{\ver}, \;
\bar{\omega}^B_{\ver, \hat{z}}=c_{\ver}-id_{\ver},  \;\; \mbox{(inter layer)}, \label{omegaAB}  
\ee
where $\hat{i}= \hx, \hy$.
It is easily verified, 
$$
\{\omega^{A\dagger}_{\ver, \hi}, \omega^B_{\ver, \hi}\}=0,
$$
etc.
By using the above notation in Eq.~(\ref{omegaAB}), the eight cube operators are schematically displayed in 
Fig~\ref{Fig2}. 
It should be remarked that it is not obvious if configurations of $\omega^A$, $\tilde{\omega}^A$, etc.
shown in Fig.~\ref{Fig2} can be constructed consistently.
We give explicit forms of the cube operators corresponding to those in Fig.~\ref{Fig2};
\be
Q^+_{\ver} = & {1\over \sqrt{8}}[-d_{\ver+\hx+\hy}+id_{\ver+\hx}+id_{\ver+\hy}-d_{\ver}  \nonumber \\
& -ic_{\ver+\hx+\hy}+c_{\ver+\hx}+c_{\ver+\hy}-ic_{\ver}],  \nonumber \\
Q^-_{\ver} = & {1\over \sqrt{8}}[-d_{\ver+\hx+\hy}+id_{\ver+\hx}-id_{\ver+\hy}+d_{\ver}  \nonumber \\
& -ic_{\ver+\hx+\hy}+c_{\ver+\hx}-c_{\ver+\hy}+ic_{\ver}],  \nonumber \\
\tilde{Q}^+_{\ver} = & {1\over \sqrt{8}}[d_{\ver+\hx+\hy}-id_{\ver+\hx}-id_{\ver+\hy}+d_{\ver}  \nonumber \\
& -ic_{\ver+\hx+\hy}+c_{\ver+\hx}+c_{\ver+\hy}-ic_{\ver}],  \nonumber \\
\tilde{Q}^-_{\ver} = & {1\over \sqrt{8}}[d_{\ver+\hx+\hy}-id_{\ver+\hx}+id_{\ver+\hy}-d_{\ver}  \label{QP} \\
& -ic_{\ver+\hx+\hy}+c_{\ver+\hx}-c_{\ver+\hy}+ic_{\ver}],  \nonumber \\
P^+_{\ver} = & {1\over \sqrt{8}}[-d_{\ver+\hx+\hy}-id_{\ver+\hx}+id_{\ver+\hy}+d_{\ver}  \nonumber \\
& -ic_{\ver+\hx+\hy}-c_{\ver+\hx}+c_{\ver+\hy}+ic_{\ver}],  \nonumber \\
P^-_{\ver} = & {1\over \sqrt{8}}[-d_{\ver+\hx+\hy}-id_{\ver+\hx}-id_{\ver+\hy}-d_{\ver}  \nonumber \\
& -ic_{\ver+\hx+\hy}-c_{\ver+\hx}-c_{\ver+\hy}-ic_{\ver}],  \nonumber \\
\tilde{P}^+_{\ver} = & {1\over \sqrt{8}}[d_{\ver+\hx+\hy}+id_{\ver+\hx}-id_{\ver+\hy}-d_{\ver}  \nonumber \\
& -ic_{\ver+\hx+\hy}-c_{\ver+\hx}+c_{\ver+\hy}+ic_{\ver}],  \nonumber \\
\tilde{P}^-_{\ver} = & {1\over \sqrt{8}}[d_{\ver+\hx+\hy}+id_{\ver+\hx}+id_{\ver+\hy}+d_{\ver}  \nonumber \\
& -ic_{\ver+\hx+\hy}-c_{\ver+\hx}-c_{\ver+\hy}-ic_{\ver}].  \nonumber
\ee
By the straightforward calculation, it is verified that all eight operators in Eq.~(\ref{QP}) 
and their hermitian conjugates \textit{located at the same cube} anti-commute with each other 
except for the commutators such as,
$\{Q^{+\dagger}_{\ver},Q^+_{\ver}\}=1$, etc.
However, some of them located at adjacent cubes do not commute with each other such as 
$\{ Q^{+\dagger}_{\ver}, {Q}^-_{\ver+\hx+\hy}\}= 1/4$, etc.
This comes from the fact that the number of sites doubles that of cubes.
Therefore, a suitable assignment of locations of the cube operators is required to construct 
the target projective Hamiltonian, which can be carried out by considering symmetry transformations.
See later discussion in Sec.~III. Here, it should be emphasized that the use of the cubic operators is essential for
constructing the bilayer models. In other words, operators defined on plaquettes cannot be introduced for it. Therefore, the bilayer models introduced in the following section cannot be transformed to systems with unconnected unit cells. See, e.g., Fig.~\ref{Fig4}.


\subsection{Time reversal symmetry of eight cube operators}

Before going to the model construction, we introduce a time-reversal symmetry (${\cal T}$) 
for the second quantized operators \cite{Ludwig} as follows, which plays an important role in later discussions;
\be
{\cal T} i {\cal T}^{-1}=-i, \;\; {\cal T} c_{\ver} {\cal T}^{-1}= d_{\ver}, \;\; {\cal T} d_{\ver}{\cal T}^{-1}=c_{\ver}.
\label{timesym}
\ee
From Eq.~(\ref{timesym}), the transformation of cube operators is induced. 
It is easily verified, ${\cal T} Q^+_{\ver} {\cal T}^{-1} =-iQ^+_{\ver}$, etc. 
The above time-reversal symmetry in Eq.~(\ref{timesym}) is reminiscent of that of the quantum spin Hall effect.
In fact, fermions $c_{\ver}$ and $d_{\ver}$ correspond to spin-up and spin-down electrons, respectively,
and ${\cal T}^2=-1$ in both systems.

LIOMs are given as the number operators of the above $\ell$-bits, $Q^+_{\ver} \sim \tilde{P}^-_{\ver}$, 
\be
&& K^+_{\ver}= Q^{+\dagger}_{\ver}Q^+_{\ver}, \;\;  K^-_{\ver}= Q^{-\dagger}_{\ver}Q^-_{\ver}, \nonumber \\
&& \tilde{K}^+_{\ver}= \tilde{Q}^{+\dagger}_{\ver}\tilde{Q}^+_{\ver}, \;\; 
 \tilde{K}^-_{\ver}= \tilde{Q}^{-\dagger}_{\ver}\tilde{Q}^-_{\ver}, \label{LIOM1} 
\ee
\be
&& M^+_{\ver}= P^{+\dagger}_{\ver}P^+_{\ver}, \;\;  M^-_{\ver}= P^{-\dagger}_{\ver}P^-_{\ver}, \nonumber \\
&& \tilde{M}^+_{\ver}= \tilde{P}^{+\dagger}_{\ver}\tilde{P}^+_{\ver}, \;\; 
 \tilde{M}^-_{\ver}= \tilde{P}^{-\dagger}_{\ver}\tilde{P}^-_{\ver}. \label{LIOM2}
\ee
All the LIOMs in Eq.~(\ref{LIOM2}) are invariant under the time-reversal transformation ${\cal T}$ in Eq.~(\ref{timesym}).
The Hamiltonians are to be constructed via the above LIOMs.
We require the target systems to have topological properties. 
In other words, on the construction of the Hamiltonian, we require to assign the system to some topological class 
in ten-fold way \cite{Altland}. 
To this end, we impose chiral symmetry on the Hamiltonians, which is
discussed in the following subsection.

\subsection{Chiral symmetries of eight cube operators}

The next step is to assign locations of these cubes to define Hamiltonian with non-trivial topology. 
It is possible to construct various models for it by means of the eight cube operators.
In the following, we show some of them.
We shall impose chiral symmetry ${\cal S}$ to the target models.
With a transformation using a unitary operator ${\cal U}$, chiral symmetry requires 
that the (second quantized) Hamiltonian $H$ transforms as~\cite{Ludwig},
\be
({\cal U} K)H ({\cal U}K)^{-1}={\cal U} H^\ast {\cal U}^{-1}=H,
\label{symm}
\ee  
where $K$ is the complex conjugation, ${\cal O}^\ast$ denotes the complex conjugate of the operator ${\cal O}$. 
The chiral operator is given by ${\cal S}={\cal U}K$, which is {\it anti-unitary operator}~\cite{Ludwig}.
The operator ${\cal S}$ plays an important role in the construction of topological models.

Here, as a candidate of the unitary operator ${\cal U}$ in ${\cal S}$,
we can introduce two unitary operators $\CS_{x}$ and $\CS_y$.
Under ${\cal S}_x$ for $\ver=(x=\mbox{even}, y)$,
\be
& c_{\ver}\to ic^\dagger_{\ver+\hx}, \; \, c_{\ver+\hx}\to -i c^\dagger_{\ver+2\hx}, \nonumber \\
& c_{\ver+\hy}\to ic^\dagger_{\ver+\hx+\hy}, \;\; c_{\ver+\hx+\hy}\to -ic^\dagger_{\ver+2\hx+\hy}, \nonumber \\
& d_{\ver}\to id^\dagger_{\ver+\hx}, \; \, d_{\ver+\hx}\to -i d^\dagger_{\ver+2\hx}, \nonumber \\
& d_{\ver+\hy}\to id^\dagger_{\ver+\hx+\hy}, \;\; d_{\ver+\hx+\hy}\to -id^\dagger_{\ver+2\hx+\hy}.
\label{trans1}
\ee
In order to construct symmetric Hamiltonians under ${\cal S}_{x}K$,
the following properties are useful,
\be 
& {\cal S}_x (Q^+_{\ver})^\ast {\cal S}^{-1}_x=-i(Q^-_{\ver+\hx})^\dagger,  \nonumber \\
& {\cal S}_x (Q^-_{\ver})^\ast {\cal S}^{-1}_x=-i(Q^+_{\ver+\hx})^\dagger,  \nonumber \\
& {\cal S}_x (\tilde{Q}^+_{\ver})^\ast {\cal S}^{-1}_x=-i(\tilde{Q}^-_{\ver+\hx})^\dagger,   \nonumber \\
& {\cal S}_x (\tilde{Q}^-_{\ver})^\ast {\cal S}^{-1}_x=-i(\tilde{Q}^+_{\ver+\hx})^\dagger,   
\label{trans2}
\ee
and similarly for $\{ P_{\ver}\}$'s. 
In the ordinary chiral symmetry, the unitary matrix ${\cal U}$ operates on internal symmetries.
In the present case, however, ${\cal S}_{x}$ operates on the site index and therefore it is a kind of 
generalized translation operator in the $x$-direction.
As a result, some specific choice of the coefficients is required in the transformation Eq.~(\ref{trans1})
coming from the built-up definition of the cube operators $Q$ and $P$ defined in Eq.~(\ref{QP}).

Transformation ${\cal S}_y$ can be defined similarly for $\ver=(x,y=\mbox{even})$, 
\be
& c_{\ver}\to -ic^\dagger_{\ver+\hy}, \; \, c_{\ver+\hx}\to -i c^\dagger_{\ver+\hx+\hy}, \nonumber \\
& c_{\ver+\hy}\to -ic^\dagger_{\ver+2\hy}, \;\; c_{\ver+\hx+\hy}\to -ic^\dagger_{\ver+\hx+2\hy}, \nonumber \\
& d_{\ver}\to id^\dagger_{\ver+\hy}, \; \, d_{\ver+\hx}\to i d^\dagger_{\ver+\hx+\hy}, \nonumber \\
& d_{\ver+\hy}\to id^\dagger_{\ver+2\hy}, \;\; d_{\ver+\hx+\hy}\to id^\dagger_{\ver+\hx+2\hy},
\label{trans3}
\ee
and under ${\cal S}_y$, 
\be 
& {\cal S}_y (Q^+_{\ver})^\ast {\cal S}^{-1}_y=-i(\tilde{Q}^+_{\ver+\hy})^\dagger,  \nonumber \\
& {\cal S}_y (Q^-_{\ver})^\ast {\cal S}^{-1}_y=-i(\tilde{Q}^-_{\ver+\hy})^\dagger,  \nonumber \\
& {\cal S}_y (\tilde{Q}^+_{\ver})^\ast {\cal S}^{-1}_y=-i({Q}^+_{\ver+\hy})^\dagger,   \nonumber \\
& {\cal S}_y (\tilde{Q}^-_{\ver})^\ast {\cal S}^{-1}_y=-i({Q}^-_{\ver+\hy})^\dagger,   
\label{trans3}
\ee
and similarly for $\{ P_{\ver}\}$'s.
The ${\cal S}_y$ is again induces a translation in the $y$-direction.
This is the origin of the slight difference between ${\cal S}_x$ and ${\cal S}_y$.

As we show in the following sections, we employ ${\cal S}_x$ as a guiding principle for
constructing model Hamiltonians.
[Chiral symmetry ${\cal S}_y$ is less effective for the construction.
Imposing both of them is incompatible for bilayer models.
See later discussion.]

\section{Models and their Hamiltonian}

\subsection{Models on two-dimensional bilayer lattice}

By using transformation properties of $\{Q\}'s$ and $\{P\}'s$ obtained in the previous section [Eq.~(\ref{trans2})], 
we can construct various models.
As we are interested in the full-localized system with a topological phase, the symmetry $S_x$ or $S_y$
can be a guiding principle for constructing Hamiltonian, which is composed of the LIOMs, i.e., $K^\pm_{\ver}$, etc in 
Eqs.~(\ref{LIOM1}) and (\ref{LIOM2}).
In order to construct models, the following facts have to be taken into account;
\renewcommand{\theenumi}{(\Alph{enumi})}
\begin{enumerate}
\item
Under the transformation in Eqs.~(\ref{trans1}), the LIOMs transform such as 
\be
K^+_{\ver} \to Q^-_{\ver +\hx}(Q^-_{\ver+\hx})^\dagger 
&=&-(Q^-_{\ver+\hx})^\dagger Q^-_{\ver +\hx}+1, \nonumber \\
&=& -K^-_{\ver+\hx}+1,
\; \mbox{etc.}
\label{additional}
\ee
\item
Commutativity of the LIOMs, $\{K_{\ver}\}$ and $\{P_{\ver}\}$ at the same location does not guarantee that
they all commute with each other.
For example, $K^+_{\ver}$ does not commute with $K^-_{\ver+\hx+\hy}$.
\end{enumerate}
From (A), signs of the LIOMs in the Hamiltonian have to be determined suitably to cancel the additional constant
in Eq.~(\ref{additional}). 
From (B), $\{K_{\ver}\}$ and $\{P_{\ver}\}$ should be call \textit{quasi-LIOMs}, although the Hamiltonian 
is to be constructed to commute with all of them.
There exits certain ``selection rule'' such that, 
$(Q^+_{\ver})^\dagger(Q^{-\dagger}_{\ver+\hx+\hy})|0\rangle$ \textit{cannot} be an eigenstate of the Hamiltonian.
Therefore, suitable assignment of spatial location of the LIOMs is required.


\begin{figure}[t]
\begin{center} 
\includegraphics[width=8.5cm]{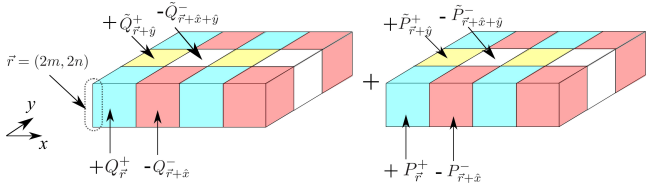} 
\end{center} 
\caption{
Schematic picture of Hamiltonian of bilayer system given by Eq.~(\ref{model1}).
Unit cell is $2\times 2$, which contains four cubes.
}
\label{Fig3}
\end{figure}

There are still various models that satisfy the above requirements coming from (A) and (B).
One of the generic ones defined on the full bilayer lattice is the following system with Hamiltonian such as,
\be
H_{\rm BL} &=& \sum_{m,n}\sum_{\ver=(2m,2n)}\Big( \tau_K[{K}^+_{\ver}-{K}^-_{\ver+\hx}
+\tilde{K}^+_{\ver+\hy}-\tilde{K}^-_{\ver+\hx+\hy}] \nonumber  \\
&&+\tau_M[{M}^+_{\ver}-{M}^-_{\ver+\hx}+\tilde{M}^+_{\ver+\hy}-\tilde{M}^-_{\ver+\hx+\hy}]\Big),
\label{model1}
\ee
where $\tau_K$ and $\tau_M$ are arbitrary real parameters, and $(n,m)$ are integers.
The spatial structure of $H_{\rm BL}$ is schematically shown in Fig.~\ref{Fig3}.
The parameters $\tau_K$ and $\tau_M$ can be site dependent as long as they satisfy the symmetry ${\cal S}_xK$, 
such as $(\tau_K, \tau_M) \to (\tau_{K,n}, \tau_{M,n})$,
but we consider the uniform case in this work.
Expression of $H_{\rm BL}$ in terms of the original fermions, $c_{\ver}$ and $d_{\ver}$, is obtained
by substituting Eqs.~(\ref{QP}), (\ref{LIOM1}) and (\ref{LIOM2}) into $H_{\rm BL}$ in Eq.~(\ref{model1}).
With the periodic boundary condition, the number of the LIOMs in $H_{\rm BL}$ is extensive,
and all energy eigenstates are localized and given as $(Q^+_{2m,2n})^\dagger|0\rangle$, etc.

Here, let us consider the symmetries of the Hamiltonian $H_{\rm BL}$. Since the model construction is carried out with
respect to the chiral symmetry ${\cal S}_xK$, the second quantized Hamiltonian $H_{\rm BL}$ is chiral symmetric.
Also, we easily notice that the Hamiltonian $H_{\rm BL}$ has the time-reversal symmetry ${\cal T}$, defined in 
the previous section. 
Furthermore, in the usual sense in the topological classification \cite{Ludwig}, the chiral
transformation ${\cal S}_xK$ is to be given by the product of the time-reversal transformation 
${\cal T}$ and a particle-hole transformation ${\cal C}$. 
Hence, we also can directly notice the presence of 
the particle-hole transformation ${\cal C}$ given as ${\cal T}^{-1} {\cal S}_xK$. 
Therefore, the Hamiltonian $H_{\rm BL}$ has ${\cal T}$, ${\cal C}$ and 
${\cal S}_xK$ symmetries. 
This fact implies that the Hamiltonian $H_{BL}$ belongs to BDI class in the ten-fold way \cite{Altland}. 
As far as the topological classification \cite{Chiu}, this fact indicates that the model $H_{\rm BL}$ can exhibit 
a topological phase and some gapless edge states for a certain lattice geometry. 
In later sections, we shall discuss such topological aspects, some of which are substantially related to the
full-localization properties.

Furthermore for the system of $H_{BL}$, nontrivial interactions can be introduced, which preserve the localization properties and chiral symmetry ${\cal S}_xK$.
One of them is given by,
\be 
 H_{\rm BLI} &=& g \sum_{m,n}\sum_{{\ver=(2m,2n)}}
\biggr[{K}^+_{\ver}{K}^-_{\ver+\hx}+\tilde{K}^+_{\ver+\hy}\tilde{K}^-_{\ver+\hx+\hy}  \nonumber \\
&& \hspace{-0.3cm}
-{K}^+_{\ver}\tilde{K}^+_{\ver+\hy}-{K}^-_{\ver+\hx}\tilde{K}^-_{\ver+\hx+\hy}
+(K \to M)\biggl],
\label{HBLI}
\ee
where $g$ is the coupling constant.
The interactions given by $H_{\rm BLI}$ mostly describe scattering processes of $c_{\ver}$ and $d_{\ver}$.
Another type of interactions can be introduced by the terms such as,
\be
H_{\rm BLII} &=& g'  \sum_{m,n}\sum_{{\ver=(2m,2n)}}
\biggl[({K}^+_{\ver}-{1\over 2})({K}^-_{\ver+\hx}-{1 \over 2})  \nonumber \\
&&+(\tilde{K}^+_{\ver +\hy}-{1 \over 2})(\tilde{K}^-_{\ver+\hx+\hy}-{1 \over 2})    \label{HBLII}   \\
&&+({K}^+_{\ver}-{1 \over 2})(\tilde{K}^+_{\ver+\hy}-{1 \over 2})  \nonumber \\
&&+({K}^-_{\ver+\hx}-{1 \over 2})(\tilde{K}^-_{\ver+\hx+\hy}-{1 \over 2})
+(K \to M)\biggl],\nonumber
\ee
which is again invariant under ${\cal S}_xK$.

If one discards chiral symmetries but preserves the integrability of models, the interactions with the following
form are possible, i.e.,
\be
H_{\rm III} = g''\sum_{m,n}\sum_{{\ver=(2m,2n)}}K^+_{\ver}P^{+\dagger}_{\ver}P^-_{\ver+\hx}+\cdots,
\ee
which are composed of $K$'s and $P$'s.
In this case, $K$'s are conserved quantities and can be fixed to certain finite values.
In the specific sector, the model reduces to a free system without genuine interaction terms.


\begin{figure}[t]
\begin{center} 
\includegraphics[width=7.5cm]{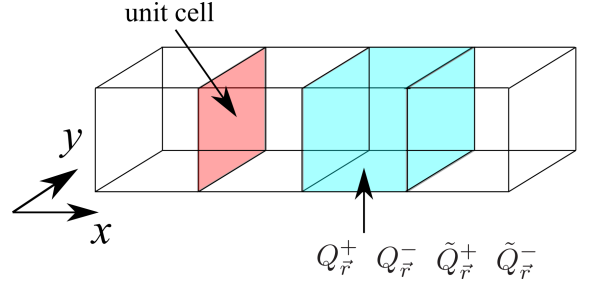} 
\end{center} 
\caption{
Schematic picture of Hamiltonian of square prime given by Eq.~(\ref{model2}).
Each cube contains four cube operators, $Q^+_{\ver} \sim \tilde{Q}^-_{\ver}$.
}
\label{Fig4}
\end{figure}

\subsection{Models on square prism lattice and topological properties}

We have shown that the model of $H_{\rm BL}$ belongs to the BDI class. 
Hence, if one reduces the model to a certain one-dimensional system without changing the symmetry class, 
the resultant (quasi-)one-dimensional model has a possibility to exhibit a topological phase characterized 
by some bulk topological invariant, e.g., winding number, Berry or Zak phase \cite{Asboth}, etc. 
From this point of view, we construct another interesting and also instructive model defined on a square prism lattice,
whose Hamiltonian is given as follows  
(see Fig.~\ref{Fig4}), 
\be
H_{\rm SP} = \sum_{\ver=(n,0)}\biggl[
\tau_0 [K^+_{\ver}-K^-_{\ver}]+\tau_1 [\tilde{K}^+_{\ver}-\tilde{K}^-_{\ver}]
\biggr],
\label{model2}
\ee
where $\tau_0$ and $\tau_1$ are arbitrary real parameters.
The model $H_{\rm SP}$ is invariant under the transformation in Eqs.~(\ref{trans1}) for $\{c_{\ver}\}$'s
with the identification $c_{\ver+2\hy}=c_{\ver}$.
The above four LIOMs per unit cube are extensive, and localized single-particle eigenstates are 
$(Q^+_{\ver})^\dagger|0\rangle, \cdots, (\tilde{Q}^-_{\ver})^\dagger|0\rangle$. 
Nontrivial interactions can be introduced as in the previous work for the Creutz ladder \cite{OKI2020}.

In later discussion, we shall study the model in Eq.~(\ref{model2}) by numerical methods.
To this end, we express the Hamiltonian, $H_{\rm SP}$, in terms of the original fermions.
After some calculation, we obtain
\be
H_{\rm SP} &=& H_{\rm CL} +H_{\rm IS1}+H_{\rm IS2},  \label{HSP2} \\
H_{\rm CL} &=& \sum_{\ver=(n,0)}\Big[
i\bar{\tau}(c^\dagger_{\ver+\hx+\hy} c_{\ver+\hy}-c^\dagger_{\ver+\hx}c_{\ver}) \nonumber \\
&&\hspace{1cm}+\bar{\tau}(c^\dagger_{\ver}c_{\ver+\hx+\hy}+c^\dagger_{\ver+\hy}c_{\ver+\hx}) \nonumber \\
&&\hspace{1cm}-i\bar{\tau}(d^\dagger_{\ver+\hx+\hy} d_{\ver+\hy}-d^\dagger_{\ver+\hx}d_{\ver}) \label{HCL} \\
&&\hspace{1cm} +\bar{\tau}(d^\dagger_{\ver}d_{\ver+\hx+\hy}+d^\dagger_{\ver+\hy}d_{\ver+\hx}) \Big]
+\mbox{h.c.},  \nonumber
\ee
\be
H_{\rm IS1} &=&  \sum_{\ver=(n,0)}\Delta
(c^\dagger_{\ver}d_{\ver+\hx}+c^\dagger_{\ver+\hx}d_{\ver} \nonumber \\
&&\hspace{0.5cm} +c^\dagger_{\ver+\hy}d_{\ver+\hx+\hy} +c^\dagger_{\ver+\hx+\hy}d_{\ver+\hy})
+\mbox{h.c.}, \label{HIS1}
\ee
\be
H_{\rm IS2} &=& \sum_{\ver=(n,0)}i\Delta
(d^\dagger_{\ver+\hx}c_{\ver+\hy}+c^\dagger_{\ver+\hx+\hy}d_{\ver}  \nonumber \\
&& \hspace{1cm} +d^\dagger_{\ver+\hy}c_{\ver+\hx}+c^\dagger_{\ver}d_{\ver+\hx+\hy})
+\mbox{h.c.},  \label{HIS2}
\ee
where $\bar{\tau}= {1 \over 4}(\tau_0+\tau_1)$ and $\Delta={1 \over 4}(\tau_1-\tau_0)$.
$H_{\rm CL}$ is nothing but the Creutz ladder Hamiltonian of 
$c_{\ver}$ and $d_{\ver}$, 
and $H_{\rm IS1}$ and $H_{\rm IS2}$ mix them and vanish for $\tau_0=\tau_1$.

\begin{figure}[t]
\begin{center} 
\includegraphics[width=8cm]{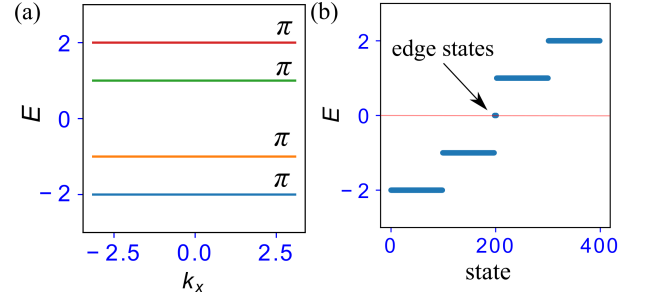} 
\end{center} 
\caption{
(a) Energy eigenvalues of $H_{\rm SP}$ for $\tau_0=2$ and $\tau_1=1$ for the periodic boundary condition.
There exist four flat-bands. 
The numbers near each band show $\gamma_{M}$, the Berry phase. 
All flat-bands are topologically nontrivial.
(b) Energy eigenvalues under the open boundary condition.
Four-fold zero-energy edge states appear.
}
\label{Fig5}
\end{figure}

Let us investigate the topological properties of the non-interacting system of $H_{\rm SP}$.
It is useful to express the Hamiltonian Eq.~(\ref{HSP2}) in terms of operators in the momentum space
for the $x$-direction; 
\be \hspace{-0.2cm}
\Phi(k_x)\equiv\Big(\tilde{c}(k_x, y=0), \tilde{c}(k_x, y=1), \tilde{d}(k_x,y=0),\tilde{d}(k_x,y=1)\Big), \nonumber
\ee
where $\tilde{c}(k_x)$'s are Fourier-transformed operators and 
$H_{\rm SP} = \int dk_x\Phi^\dagger(k_x)h_{\rm SP}(k_x)\Phi(k_x)$.
Explicitly, $h_{\rm SP}(k_x)$ is given as,
\be
\hspace{-0.5cm}
h_{\rm SP}(k_x) = \begin{pmatrix}
                       2\bar{\tau}s(k)  & 2\bar{\tau}c(k)  & 2\Delta c(k) & 2\Delta s(k)  \\
                       2\bar{\tau}c(k)  & -2\bar{\tau}s(k)  & -2\Delta s(k) & 2\Delta c(k)  \\
                       2\Delta c(k) & -2\Delta s(k)  &   -2\bar{\tau}s(k)  & 2\bar{\tau}c(k)  \\
                       2\Delta s(k) & 2\Delta c(k)  &  2\bar{\tau}c(k)  & 2\bar{\tau}s(k)
                      \end{pmatrix}, \nonumber\\
                      \label{hSP}
\ee
where $s(k)\equiv \sin (k_x)$ and $c(k)\equiv \cos (k_x)$.

From Eq.~(\ref{hSP}), the symmetries of $h_{\rm SP}(k_x)$ is clear [besides ${\cal S}_x$ in Eq.~(\ref{trans1})]. 
First, $h_{\rm SP}(k_x)$ has the time-reversal symmetry mentioned in Sec.~II.B, 
\be
&T_b h_{\rm SP}(k_x) T_b^{-1} =  h_{\rm SP}(-k_x),  \nonumber  \\
&T_b = K\begin{pmatrix}
                0 & {\bf 1}_2 \\
                {\bf 1}_2 & 0
                \end{pmatrix},
\label{inversionsym1}
\ee
where $K$ is complex conjugate operator and ${\bf 1}_2$ is $2\times 2$ identity matrix. 
Hence, $T_b$ is anti-unitary.

Second, $h_{\rm SP}(k_x)$ has a particle-hole symmetry, 
\be
&C_b h_{\rm SP}(k_x) C_b^{-1} =  -h_{\rm SP}(-k_x),  \nonumber  \\
&C_b = \begin{pmatrix}
                0 & -i\sigma_y \\
                i\sigma_y & 0
                \end{pmatrix},
\label{inversionsym2}
\ee
where $\sigma_y$ is the $y$-component of Pauli matrix. 
Hence, $C_b$ is unitary.

Third, $h_{\rm SP}(k_x)$ has a chiral symmetry given by the chiral operator $S_b=T_b C_b$, 
\be
&S_b h_{\rm SP}(k_x) S_b^{-1} =  -h_{\rm SP}(k_x),  \nonumber  \\
&S_b = K\begin{pmatrix}
                \sigma_y & 0 \\
                0 & -\sigma_y
                \end{pmatrix},
\label{inversionsym3}
\ee
where $S_b$ is anti-unitary.
From these symmetries, $h_{\rm SP}(k_x)$ belongs to the BDI class in ten-fold way \cite{Altland,Chiu}. 

Furthermore, the system $H_{\rm SP}$ also has a spatial reflection (inversion) symmetry, which is given by
\be
&I h_{\rm SP}(k_x) I^{-1} =  h_{\rm SP}(-k_x),  \nonumber  \\
&I = \begin{pmatrix}
                \sigma_x & 0 \\
                 0 & \sigma_x
                \end{pmatrix},
\label{inversionsym4}
\ee
where $\sigma_x$ is the $x$-component Pauli matrix.
This reflection symmetry plays an important role for the quantization of Berry (Zak) phase \cite{Zak1,Zak2}, 
which acts as a topological index in this system.

We numerically demonstrate the topological properties of $h_{\rm SP}(k_x)$. 
By diagonalizing $h_{\rm SP}(k_x)$, we can obtain energy eigenvalues as shown in Fig.~\ref{Fig5} (a). 
Certainly, there appear four flat-bands. 
Here, we calculate Berry phase \cite{Asboth} given by 
$\gamma_{M}=i\int^{\pi}_{-\pi}\langle u_{\ell}(k_x)|\partial_{k_x}|u_{\ell}(k_x)\rangle \; dk_x$, 
where $|u_{\ell} (k_{x})\rangle$ is $\ell$-th eigenstate of $h_{\rm SP}(k_x)$. For each flat-bands, 
$\gamma_M$ takes $\pi$, that is, each bands are non-trivial. 
This quantization comes from the inversion symmetry $I$ (crystalline topological insulator \cite{Fang}). 
We shall show the another Berry phase obtained by introducing boundary twist, in later section.

Here, we also show the energy spectrum by diagonalizing $H_{\rm SP}$ under the OBC .
The result indicates the existence of the gapless edge modes, which are discussed in Sec.~IV.

We introduce the following ``inter-leg" hopping, which respects the symmetries of the Hamiltonian $H_{SP}$.
in 
Eq.~(\ref{trans1}),
\be
H_v=v \sum_{\ver=(n,0)}(c^\dagger_{\ver+\hy}c_{\ver}+d^\dagger_{\ver+\hy}d_{\ver}+\mbox{h.c.}),
\label{Hv}
\ee
where $v$ is an arbitrary real parameter.
It should be noted that Eq.~(\ref{Hv}) does not break the chiral and reflection symmetries.
In the previous works on the Creutz ladder, we showed that the hopping $H_v$ makes all states extended 
even for infinitesimal $v$.
However, needless to say,
the symmetries for topological properties are preserved, thus bulk-band  topology does 
not change without gap closing. 
Also, we expect that the corresponding gapless edge modes are preserved even for finite $v$. 

It is interesting and also important to examine if there exist interactions that respect 
the ${\cal S}_x$-symmetry [Eq.~(\ref{trans1})].
To search them, Eq.~(\ref{additional}) is quite useful.
There are several forms of the interactions, and we display typical one;
\be
H_{\rm SPI} = &&\lambda\sum_{\ver}\Big[(K^+_{\ver}-{1 \over 2})(K^+_{\ver+\hx}-{1 \over 2})  \nonumber \\
&&\hspace{1cm} +(K^-_{\ver}-{1 \over 2})(K^-_{\ver+\hx}-{1 \over 2})   \nonumber \\
&&\hspace{1cm}+(\tilde{K}^+_{\ver}-{1 \over 2})(\tilde{K}^+_{\ver+\hx}-{1 \over 2})   \nonumber \\
&&\hspace{1cm}+(\tilde{K}^-_{\ver}-{1 \over 2})(\tilde{K}^-_{\ver+\hx}-{1 \over 2})\Big],
\label{HSPI}
\ee
where $\lambda$ is coupling constant.
Here, it should be noted that the above interaction $H_{\rm SPI}$ [Eq.~(\ref{HSPI})] is invariant
under transformations Eqs.~(\ref{inversionsym1}) $\sim$ (\ref{inversionsym4}) as well as ${\cal S}_xK$.
Therefore, we expect that
the interaction $H_{\rm SPI}$ plays an important role for the emergence of gapless edge modes
as verified later on.
There, we shall also explain that a phase transition takes place as the parameter $\lambda$ is varied.
Discussion on topological properties of the system will be given as well.

In later numerical study on the Hamiltonian $H_{SP}$, we study effects of the following interactions, which
are invariant under the transformations Eqs.~(\ref{inversionsym1}) $\sim$ (\ref{inversionsym4}), 
\be
\hspace{-0.5cm}
H_{\rm II} &=& V \sum_{\ver}\biggr[n^c_{\ver}n^c_{\ver+\hy}+
n^c_{\ver}n^c_{\ver+\hx}+n^c_{\ver+\hy}n^c_{\ver+\hx+\hy}\nonumber\\
&&+ (c\to d)\biggl],
\label{HII}
\ee
where $V$ is an arbitrary real parameter, and 
$n^c_{\ver}=c^\dagger_{\ver}c_{\ver}$, $n^d_{\ver}=d^\dagger_{\ver}d_{\ver}$.
The interactions $H_{\rm II}$ in Eq.~(\ref{HII}) seem to break the ${\cal S}_x$-symmetry in Eq.~(\ref{trans1}).
In fact under Eq.~(\ref{trans1}), there appear terms such as 
$\sum_{\ver}[n^c_{\ver}+n^c_{\ver+\hx}+n^c_{\ver+\hy}+n^c_{\ver+\hx+\hy}+ (c\to d)]$,
in addition to a constant.
However as we always consider the system with fixed particle number, these terms are irrelevant.
Furthermore, the additional constant in the Hamiltonian does not change wave functions of energy eigenstates,
it is also irrelevant.
Therefore, we expect the stability of topological properties of the Hamiltonian $H_{SP}$ in the presence of 
$H_{\rm II}$, as long as it does not change the band structure.
This expectation will be verified by the numerical calculation later on.

\section{Edge modes}


\begin{figure}[t]
\begin{center} 
\includegraphics[width=6cm]{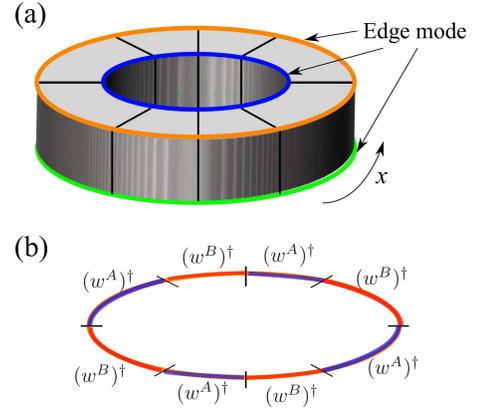} 
\end{center} 
\caption{
(a) Coarse grained schematic picture of bilayer system with open boundary condition
of thin cylinder (disk) shape.
(b) There emerge four edge modes, each of which is one-dimensional and a linear combination
of $(\omega^A)^\dagger$'s \textit{or}  $(\omega^B)^\dagger$'s.
}
\label{Fig6}
\end{figure}

In the previous section, we have introduced models $H_{\rm BL}$ and $H_{\rm SP}$ 
of full localization in the bulk and belong to BDI class.
Then, it is expected that there appear gapless edge modes in the above models in the OBC.
Depending on the geometrical structure of the systems, gapless edge modes emerge in a different way.
Then, we shall discuss the models $H_{\rm BL}$ and $H_{SP}$, separately.

Let us first consider the model $H_{\rm BL}$ in a thin cylinder lattice whose
schematic picture is displayed in Fig.~\ref{Fig6}.
In the $x$-direction, the system is periodic, whereas in the $y$-direction, the boundaries exist.
One may expect that gapless edge modes appear in the boundary surfaces, but this is not the case.
They exist in the four edges of the cylinder [see Fig.~\ref{Fig6} (a)].

From Fig.~\ref{Fig3}, these edges of the cylinder are composed of edges
of a sequence of $\{K\}$'s such as $(\cdots K^+K^-K^+K^-\cdots)$ or 
$(\cdots \tilde{K}^+\tilde{K}^-\tilde{K}^+\tilde{K}^-\cdots)$
[and also sequences of $\{M\}$'s].
Then from Fig.~\ref{Fig2}, the terms in the Hamiltonian corresponding to 
the boundaries of the upper plane is given by 
$$
h_{\rm B}=\tau_{K(M)}\sum'_{\ver\in {\cal C}} [(\omega^{A\dagger}_{\ver,\hx}\omega^A_{\ver,\hx})
-
(\omega^{B\dagger}_{\ver+\hx,\hx}\omega^B_{\ver+\hx,\hx})]+\cdots,
$$where ${\cal C}$ represents
one of the two closed edges in the upper plane, and $\sum'$ denotes 
the restricted summation such as, 
$\ver=(x=\mbox{even},y=\mbox{location of edge})$.] 
[Similar boundary Hamiltonian of the lower edge is obtained by the time-reversal transformation,
and then $\omega^{A/B} \to \tilde{\omega}^{A/B}$.]
The boundary Hamiltonian $h_{\rm B}$ does not contain $\omega^B_{\ver,\hx}$
and $\omega^A_{\ver+\hx,\hx}$. 
Therefore, the zero modes are created by the operators satisfying the following equation (see Fig.~\ref{Fig6} (b)),
\be
{\cal B}_{\cal C}&=&\sum'_{\ver\in {\cal C}} \alpha_{\ver}\omega^{B\dagger}_{\ver,\hx} 
=\sum'_{\ver\in {\cal C}} \alpha_{\ver+\hx}\omega^{A\dagger}_{\ver+\hx,\hx},
\label{BC}
\ee
where $\{\alpha_{\ver}\}$'s are suitably chosen phase factors as 
${\cal B}_{\cal C}$ commutes with 
the boundary Hamiltonian $h_B$.
[The commutativity between ${\cal B}_{\cal C}$ and the other parts of 
the Hamiltonian $H_{\rm SP}$ is obvious.]
It should be remarked that this commutativity is preserved even in the presence of the interactions
$H_{\rm BLI}$ and $H_{\rm BLII}$ in Eqs.~(\ref{HBLI}) and (\ref{HBLII}).
The above operator on ${\cal C}$, ${\cal B}_{\cal C}$, is a linear combination of $c^\dagger_{\ver}$
 with suitably chosen coefficients.
For example for a square edge with four sites $(1, \cdots, 4)$, we have 
${\cal B}^\dagger_{\cal C}=c^\dagger_1+ic^\dagger_2+c^\dagger_3+ic^\dagger_4
=(c^\dagger_1+ic^\dagger_2)+(c^\dagger_3+ic^\dagger_4)
=i(c^\dagger_2-ic^\dagger_3)+i(c^\dagger_4-ic^\dagger_1)$
[See Eq.~(\ref{BC})].
It is obvious that in general, ${\cal B}_{\cal C}$ can be constructed consistently with the condition Eq.~(\ref{BC}) 
because ${\cal C}$ is composed of an even number of sites.
It is also verified that ${\cal B}_{\cal C}$ transforms suitably under chiral symmetry -transformation (${\cal S}_xK$ in Eq.~(\ref{trans1}));
\be
{\cal S}_x ({\cal B^\dagger)}^\ast_{\cal C}{\cal S}^{-1}_x &=&
{\cal S}_x(c^\dagger_1-ic^\dagger_2+c^\dagger_3-ic^\dagger_4){\cal S}^{-1}_x  \nonumber  \\
&=&-ic_2+c_3-ic_4+c_1  \nonumber \\
&=&(c^\dagger_1+ic^\dagger_2+c^\dagger_3+ic^\dagger_4)^\dagger  \nonumber \\
&=&{\cal B}_{\cal C}.
\label{chiral}
\ee
Origin of this \textit{one-dimensional gapless edge modes} is closely related to flat-band localization,
and will be discussed in detail in Sec.~VI, after examining the model of $H_{SP}$. 
Here, we should note that seen from the form of Eq.~(28) the above edge zero modes survive under the additional open boundary
condition in the $x$-direction even though the ${\cal S}_xK$ symmetry is explicitly broken at the edges of the $x$-direction.

A few comments are in order.
There are four boundary modes, $\CB^a_{\cal C} \; (a=1,2,3,4)$, where the suffix $a$ denotes the four edges
of the cylinder.
In the previous works~\cite{Bahri},
it was argued that gapless edge modes are stable \textit{even at finite temperature} 
if all the other bulk states are localized.
The present bilayer model has such properties even though the gapless edge modes are
linear combination of the original particles.

As discussed in Ref.~\cite{You2017} for one-dimensional models, SYK-type models can be constructed 
via ${\cal B}_{\cal C}$ 's.
There, fourth-order terms of them are leading because of chiral symmetry given in Eq.~(\ref{chiral}). 
In the original work of the SYK model, this symmetry was imposed by hand, but it emerges naturally
in the present system.
Possible effective Hamiltonian is such that $H=\sum V_{abcd}\CB^{a\dagger}\CB^{b\dagger}\CB^c\CB^d$, where
the edge index $(a\sim d)=(1\sim 4$) and the coefficients $V_{abcd}$ are complex random numbers.
The classification of the above model \textit{for ordinary complex particles} has been already done in 
Ref.~\cite{You2017}.

Next, let us consider the model $H_{\rm SP}$ in Eq.~(\ref{model2}) with the OBC 
such as $0\le x \le L-1$.
There are four edge modes,
which are given by $w^A_{x=0,\hy}|0\rangle, \tilde{w}^B_{x=0,\hy}|0\rangle, w^B_{x=L,\hy}|0\rangle$
and $\tilde{w}^A_{x=L,\hy}|0\rangle$ [See Fig.~\ref{Fig5} (b).].
Even in the presence of the interactions $H_{\rm SPI}$ in Eq.~(\ref{HSPI}),
the four operators $(w^A_{x=0,\hy}, \cdots, \tilde{w}^A_{x=L,\hy})$ commute exactly with the Hamiltonian
$H_{\rm SP}+H_{\rm SPI}$. 
This indicates that the gapless edge modes survive in the interacting system.


\begin{figure}[t]
\begin{center} 
\includegraphics[width=6.5cm]{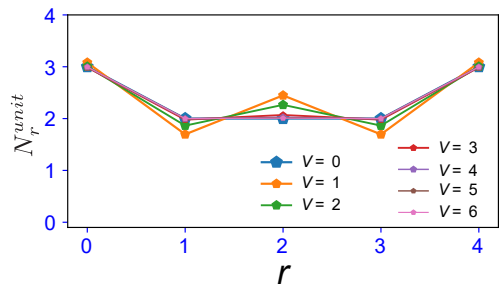} 
\end{center} 
\caption{
Density profile of half-filled + two particle for various interaction strength $V$. 
We plot the total density of the unit cell, $N^{unit}_r=n^{c}_{r+y}+n^{c}_{r}+n^{d}_{r+y}+n^{d}_{r}$.
Edge modes are quite stable against to the repulsion. 
The system size is $L=5$ (20 sites) with $12$ particles.
}
\label{Fig7}
\end{figure}

We also study effects of the interactions $H_{\rm II}$ [Eq.~(\ref{HII})]
by numerical methods \cite{Quspin}.
In particular as the above edge-mode creation operators do not commute with $H_{\rm II}$,
we are interested in stability of the edge modes.
In Fig.~\ref{Fig7}, we display the density profiles for various values of $V$ for the half-filled + two particles.
The additional two particles on top of the half-filled state are expected to correspond to the gapless edge modes.
Numerical calculations obviously show the stability of the edge modes even for large $V$.
We think that this result comes from the fact that $H_{\rm II}$ preserves symmetries, as we discussed in the above,
and then it enhances homogeneity of the bulk regime.

In the following subsection, we shall investigate topological indexes corresponding to the model 
$H_{\rm SP}+H_{\rm SPI}$
in Eqs.~(\ref{model2}) and (\ref{HSPI}).

\section{Bulk topological indexes and string operator}

In the previous section, we found that the gapless edge modes emerge under the OBC in the model $H_{\rm SP}$.
This fact implies that the present systems include SPT phases.
We further investigate topological indexes characterizing topological properties of the model $H_{\rm SP}$
in Eq.~(\ref{model2}), i.e., Berry phases obtained from a local twist, string order, etc. 

In the momentum representation, we mention that
the Berry phase in the $H_{\rm SP}$ system, $\gamma_{M}$, are quantized because of the reflection symmetry in Eq.~(\ref{inversionsym4}), and take $\gamma_{M}=0, \pi \; (\mbox{mod} \; 2\pi)$.
Obtained results of $\gamma_M$ are shown in Fig.~\ref{Fig5}.
Here, we employ another method for calculating Berry phase, which can be used for interacting systems.

To this end, we introduce local twist with $\theta \in$ ($S^{1}$: $[0,2\pi$)) for {\it all} the hopping terms in
$H_{\rm SP}$ residing on certain unit cells.
Under this local twist, the Hamiltonian $H_{\rm SP}$ depends on $\theta$, that is, $H_{\rm SP}(\theta)$. 
For the Hamiltonian $H_{\rm SP}(\theta)$, if the ground state is unique and 
gapped for all $\theta$, 
then the $Z_2$-Berry phase \cite{Hatsugai2005, Hatsugai2006,Hatsugai2007,Katsura2007,Hirano2008,HM2011} from the local twist is given by 
\be
\gamma_{L}=i\int^{2\pi}_{0}d\theta \langle g(\theta)|\partial_{\theta}|g(\theta)\rangle, 
\ee
where $|g(\theta)\rangle$ is the gapped unique ground state for $H_{\rm SP}(\theta)$. Here, we should note that the Berry phase is 
defined mod $2\pi$~\cite{Hatsugai2006}.
In the present system, the Berry phase $\gamma_{L}$ can be analytically treated 
since the exact many-body ground state of 
$H_{\rm SP}$ is already known.
The ground state is also useful for the study on the interacting case with $H_{\rm SPI}$ and 
$H_{\rm II}$ in Eqs.~(\ref{HSPI}) and (\ref{HII}).

To calculate Berry phases practically, we introduce the following twisted hopping in the cube operators at site 
$\ver=\verz=(0,0)$;
\be
Q^+_{\verz}(\theta) = & {1\over \sqrt{8}}[-\IT d_{\verz+\hx+\hy}+i\IT d_{\verz+\hx}+id_{\verz+\hy}-d_{\verz} 
 \nonumber \\
& -i\IT c_{\verz+\hx+\hy}+\IT c_{\verz+\hx}+c_{\verz+\hy}-ic_{\verz}],
\ee
and similarly for $Q^-_{\verz}(\theta), \tilde{Q}^+_{\verz}(\theta)$ and $ \tilde{Q}^-_{\verz}(\theta)$. 
In fact in the Hamiltonian with the twist $H(\theta)$, the hopping terms 
in the $x$-direction are changed to 
$K^+_{\verz}(\theta)\equiv
(Q^+_{\verz}(\theta))^\dagger Q^+_{\verz}(\theta)
\propto e^{-i\theta}d^\dagger_{\verz+\hx+\hy}d_{\verz+\hy}+\cdots.
$
It is easy to verify that the above twisted operators satisfy the same 
commutation relations with the operators for $\theta=0$, i.e.,
$$
\{(Q^+_{\verz}(\theta))^\dagger, Q^+_{\verz}(\theta)\}=1, \;
\{(Q^+_{\verz}(\theta))^\dagger, Q^+_{\ver}\}=0,
$$
for $\ver\neq \verz$, etc. 
Then, operators $K^+_{\verz}(\theta)$, etc, which are composed of $Q^+_{\verz}(\theta)$, etc, are LIOMs, and
energy eigenstates are given by $Q^{+\dagger}_{\verz}(\theta)|0\rangle$, etc.

As an example, we first consider the grounstate at $1/4$-filling for $\tau_0>\tau_1$, whose wave function
is given by,
\be
|\psi_1(\theta)\rangle = [Q^-_{\verz}(\theta)]^\dagger\prod_{\ver\neq \verz}
Q^{-\dagger}_{\ver}|0\rangle.
\label{twistGS1}
\ee
It is easily verified that the energy gap between $|\psi_1(\theta)\rangle$ and 
the excited states
does not close for any $\theta \in [0,2\pi]$.
Then, 
\be
i\gamma_{L}&=& \int^{2\pi}_0d\theta \langle \psi_1(\theta)|\partial_{\theta} 
|\psi_1(\theta)\rangle \nonumber  \\
&=& i {1\over 8}\int^{2\pi}_0 d\theta \langle 0|[-\IT d_{\verz+\hx+\hy}+i\IT d_{\verz+\hx}  \nonumber \\
&&\hspace{2cm} -i\IT c_{\verz+\hx+\hy}+\IT c_{\verz+\hx}]^\dagger  \nonumber \\
&&\times [-\IT d_{\verz+\hx+\hy}+i\IT d_{\verz+\hx}
 -i\IT c_{\verz+\hx+\hy}+\IT c_{\verz+\hx}]|0\rangle   \nonumber  \\
&=& i\pi.
\label{berry1}
\ee
Therefore, $\gamma_{L}=\pi$ for the ground state at $1/4$-filling.
The same result has been obtained by using the momentum representation of the Hamiltonian $H_{\rm SP}$~(see Fig.~\ref{Fig5} (a))
in Sec.~III.B.
On the other hand for the case $\tau_1 > \tau_0$, the ground state is given as
\be
|\psi_2(\theta)\rangle = [\tilde{Q}^-_{\verz}(\theta)]^\dagger\prod_{\ver\neq \verz}
\tilde{Q}^{-\dagger}_{\ver}|0\rangle.
\label{twistGS2}
\ee
Similar calculation to the above shows that Berry phase of $|\psi_2(\theta)\rangle$ is $\gamma_{L}=\pi$.
Transition between $|\psi_1(\theta)\rangle$ and $|\psi_2(\theta)\rangle$ takes place at $\tau_0=\tau_1$.
At this transition point  $\Delta=0$, the system is simply two independent Creutz ladder fermions, and
there appears tremendous degeneracy. 
As a result, Berry phase cannot be defined properly.

Let us turn to the half filling case.
The ground state wave function is given as 
\be
|\psi_3(\theta)\rangle = [{Q}^-_{\verz}(\theta)]^\dagger[\tilde{Q}^-_{\verz}(\theta)]^\dagger
\prod_{\ver\neq \verz}{Q}^{-\dagger}_{\ver}
\tilde{Q}^{-\dagger}_{\ver}|0\rangle.
\label{twistGS3}
\ee
Berry phase is calculated as 
\be
i\gamma_{L} &=& \int^{2\pi}_0d\theta \langle 0|[{Q}^-_{\verz}(\theta)][\tilde{Q}^-_{\verz}(\theta)]
\partial_{\theta}\Big[[{Q}^-_{\verz}(\theta)]^\dagger[\tilde{Q}^-_{\verz}(\theta)]^\dagger
\Big]|0\rangle
\nonumber \\
&=&{1 \over 16}\int^{2\pi}_0d\theta \langle 0|[D(\theta)C(\theta)]
\partial_{\theta}\Big[[C(\theta)]^\dagger[D(\theta)]^\dagger
\Big]|0\rangle   \nonumber \\
&=&{1 \over 4}\int^{2\pi}_0d\theta \langle 0|\Big[ C(\theta)\partial_\theta C^\dagger(\theta)
+D(\theta)\partial_\theta D^\dagger(\theta)\Big]|0\rangle   \nonumber \\
&=& i2\pi \nonumber \\
&=& 0, \;\; (\mbox{mod} \; 2\pi),
\label{berry2}
\ee
where $C(\theta)= -i\IT c_{\ver+\hx+\hy}+\IT c_{\ver+\hx}-c_{\ver+\hy}+ic_{\ver}$,
and $D(\theta)=\IT d_{\ver+\hx+\hy}-i\IT d_{\ver+\hx}+id_{\ver+\hy}-d_{\ver} $.
The above calculation shows that $Q^-$ and $\tilde{Q}^-$--sectors
(and $c$ and $d$-particles) contribute to Berry phase additively.
Berry phases of other states including $Q^{+\dagger}_{\ver}$, etc are calculated similarly,
and similar results are obtained.

\begin{figure*}[t]
\begin{center} 
\includegraphics[width=16cm]{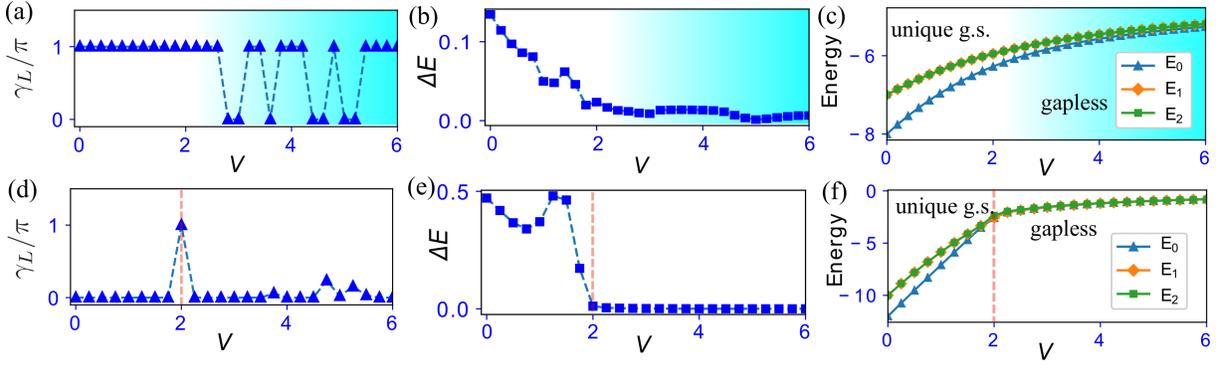} 
\end{center} 
\caption{
(a) Berry phase $\gamma_{L}$ as a function of $V$ for $1/4$-filling.  
For $V>V_c \simeq 2.7$, the Berry phase is unstable.
(b) The energy gap $\Delta E$ for $1/4$-filling 
(the difference between ground state and first excited state.) 
Both results indicate the existence of some kind of crossover in the vicinity of $V_c$.
(c) Energy of the ground state, first and second excited states without twist.
The energy gap of the ground state decreases as $V$ increases, but it keeps finite even for $V>V_c$.
System size $L=4$ ($4 \times 4$ sites) and 4 particles.
(d) Berry phase $\gamma_{L}$ as a function of $V$ for half-filling. For $V>V_c \simeq 2$, the Berry phase is unstable.
(e) The energy gap $\Delta E$ for half-filling 
(the difference between ground state and first excited state.) 
Both results indicate the existence of a phase transition in the vicinity of $V_c=2$.
(f) Energy of the ground state, first and second excited states without twist for half-filling. 
The gapless ground state emerges for $V>V'_c\simeq 2.0$.
System size $L=4$ ($4 \times 4$ sites) and 8 particles.}
\label{Fig8}
\end{figure*}

Let us consider the effects of the interactions $H_{\rm SPI}$ in Eq.~(\ref{HSPI}), which preserve
chiral symmetries and the gapless edge modes.
As we mentioned in the above, `phase transition' between two states 
$|\Psi_1\rangle=\prod_{\ver}(Q^{-\dagger}_{\ver})|0\rangle$ and 
$|\Psi_2\rangle=\prod_{\ver}(\tilde{Q}^{-\dagger}_{\ver})|0\rangle$ takes place as varying
values of $\tau_0$ and $\tau_1$.
At 1/4-filling for $\tau_0>\tau_1>2\lambda(>0)$, the ground state is given by $|\Psi_1\rangle$,
and the Berry phase $\gamma_{L}=\pi$ as we calculated.
As $\lambda$ increases, the intra-species NN repulsions getting stronger, and at $2\lambda=\tau_0-\tau_1$,
there emerge tremendous degeneracies, i.e., states such as  $|\Psi_1\rangle$, 
$\prod(\cdots, Q^{-\dagger}_{\ver-\hx}\tilde{Q}^{-\dagger}_{\ver}Q^{-\dagger}_{\ver+\hx}\cdots|0\rangle$,
$\prod(\cdots, Q^{-\dagger}_{\ver-\hx}\tilde{Q}^{-\dagger}_{\ver+\hx}Q^{-\dagger}_{\ver+\hx}\cdots|0\rangle$,
etc. have all the same energy.
Because of this degeneracy, the Berry phase is undefined. 
Even in this case, the gapless edge modes exist but they cannot be identified because of the tremendous degeneracy.
As the intra-species NN repulsion is getting stronger 
[$2\lambda>\tau_0-\tau_1$],  all cubes in one subsystem, say even cubes,
are occupied by $Q^-_{\ver}$, whereas odd cubes are empty or occupied by $\tilde{Q}^-_{\ver}$.
The number of the empty cubes is equal to that of the doubly-occupied cubes as there is no inter-species repulsion between $Q^-_{\ver}$ and $\tilde{Q}^-_{\ver}$. 
When we further add on-cube inter-species repulsion such as 
$\lambda'\sum (K^-_{\ver}-{1 \over 2})(\tilde{K}^-_{\ver}-{1 \over 2})$,
the degeneracy is resolved.
The ground state is simply doubly-degenerate and has a N\`{e}el-type order, i.e., 
$(\cdots Q^{-\dagger}_{\ver-\hx}\tilde{Q}^{-\dagger}_{\ver}Q^{-\dagger}_{\ver+\hx}
\tilde{Q}^{-\dagger}_{\ver+2\hx}\cdots)|0\rangle$.

In the thermodynamic limit, these two states are totally disconnected, and 
Berry phase can be defined for each state as in the ordinary local order
parameter such as magnetization in the Ising model.
Each state has Berry phase $\gamma_{L}=\pi$. 

In the following, we shall calculate Berry phases $\gamma_{L}$ for the system with the additional 
interaction, $H_{\rm II}$ [Eq.~(\ref{HII})], by numerical methods. 
The form of interaction $H_{\rm II}$ does not break the inversion symmetry of Eq.~(\ref{inversionsym4}).  
In general, without gapclosing, the topological phase is robust for the presence of $H_{\rm II}$.
The interaction does not allow each particle to become the CLSs of $Q^{+}$, $Q^{-}$, $\tilde{Q}^{+}$ and $\tilde{Q}^{-}$ since the number operators of the CLSs no longer commute the total Hamiltonian $H_{SP}+H_{\rm II}$. The CLSs are deformed by the interactions. Along with this, by varying the value of $V$, the many-body gap can vanish. 
Hence, the interaction $H_{\rm II}$ can induce a topological phase transition.

In particular, we are interested in how states change as $V$ is increased.
In order to have a well-defined Berry phase, the energy gap $\Delta E(\theta)$ between the ground state
and first excited states has to be positive for any $\theta \in [0,2\pi]$.
Then, we define $\Delta E \equiv \mbox{Min} [\Delta E(\theta)]$, and calculate it numerically.
The results of $\gamma_L$ and $\Delta E$ are shown in Fig.~\ref{Fig8} (a) and (b) for the $1/4$-filling state.
Data show that the Berry phase $\gamma_L=\pi$ for $V<V_c\simeq 2.7$ indicating that the state is topologically
non-trivial, whereas from $V=V_c$ the Berry phase begins to be unstable and random.
$\Delta E$ is also vanishingly small for $V>V_c$.
Similar behavior of Berry phase was observed in Ref.~\cite{Guo}.
In Fig.~\ref{Fig8} (c), we also show energies of the ground state, first and second excited states without twist.
The energy gap of the ground state decreases and gradually vanishes as a function of $V$. This indicates that the system turns into a gapless metallic phase although the critical transition point cannot be extracted due to the finite size effects.
On the other hand, we also calculate the half-filled case, 
the results of $\gamma_{L}$, $\Delta E$ and energies of the ground state, first and second excited states without twist are shown in Fig.~\ref{Fig8} (d)-(f). 
The Berry phase $\gamma_{L}$ in Fig.~\ref{Fig8} (d) is stable and stays zero for $V\lesssim 2$. For $V\gtrsim 2$ it turns into unstable and random since $\Delta E$ vanishes as shown in Fig.~\ref{Fig8} (e). 
The calculations of energies in Fig.~\ref{Fig8} (f) show that the exists for $V>V'_c\simeq 2.0$, and apparently a gapless metallic state emerges for $V>V'_c$. The difference between the $1/4$-filling and half-filling states comes from the fact that the topological non-trivial state with $\gamma_L=\pi$ is realized at the $1/4$ filling, whereas the trivial state with $\gamma_L=0$ (mod $2\pi$) at half filling.

In the above, we investigated the local quantity, Berry phase, related to topological properties of the systems.
As shown in the previous work on the Creutz ladder~\cite{OKI2020}, there is a nonlocal order parameter of
$Z_2$-topological symmetry, i.e., the string operator \cite{Kitaev,Fendley,McGinley}.
We can define a similar quantity in the present bilayer systems, which we call $Z_2$-order parameter
and string operator.
These operators are defined in terms of the LIOMs, and for the Hamiltonian $H_{SP}$,
\be
Z_2=(-1)^{\sum_{\ver}(K^+_{\ver}+K^-_{\ver}+\tilde{K}^+_{\ver}+\tilde{K}^-_{\ver})}.
\label{Z2}
\ee
As $K^-_{\ver}=Q^{-\dagger}_{\ver}Q^-_{\ver}$, etc, $Z_2$-operator in Eq.~(\ref{Z2})
is closely related to the Berry phase calculated in the above.
Under the OBC considered in the above, the $Z_2$ operator tends to 
$$
Z_2 \to (-1)^{\sum^L_{n=0}\sum_{\ver=(n,0)}
(c^\dagger_{\ver}c_{\ver}+c^\dagger_{\ver+\hy}c_{\ver+\hy}+d^\dagger_{\ver}d_{\ver}
+d^\dagger_{\ver+\hy}d_{\ver+\hy})},
$$
where we have added the terms such as 
$[\omega^{B\dagger}_{(0,0),\hy}\omega^B_{(0,0),\hy} 
+\tilde{\omega}^{A\dagger}_{(L,0),\hy}\tilde{\omega}^A_{(L,0),\hy}]$ to make $Z_2=\pm 1$.
On the other hand for the string operator, ${\cal O}(\ell,m)$, we define it as follows;
\be
{\cal O}(\ell,m)= (-1)^{\sum^m_{n=\ell}\sum_{\ver=(n,0)}
(K^+_{\ver}+K^-_{\ver}+\tilde{K}^+_{\ver}+\tilde{K}^-_{\ver})}.
\label{string}
\ee

In the above, we studied the `phase transition' caused by the interactions.
For small $\lambda$, the state has the Berry phase $\gamma_{L}=\pi$.
In this state $|\Psi_1\rangle$, $\langle K^-_{\ver}\rangle=1$ and the other expectation values are vanishing, 
and therefore $\langle {\cal O}(\ell,m)\rangle \neq 0$.
On the other hand for $2\lambda \ge \tau_0-\tau_1$, there exist large number of degenerate states, in which
 $\langle K^-_{\ver}\rangle=0$ or $1$, randomly.
Therefore, the micro-canonical ensemble gives $\langle {\cal O}(\ell,m)\rangle=0$.
By adding the on-cube inter-species repulsion, the degeneracy is resolved except the macroscopic one, and 
then $\langle {\cal O}(\ell,m)\rangle \neq 0$.
The above consideration of the various states shows that the Berry phase and the string operator
give the consistent results as topological order parameters.

Finally, we study the $1/4$-filling state by using partial-reflection (PR) overlap of wave functions,
which is recently proposed for detailed investigation of the SPT phase~\cite{PollTur,Shapourian1}. 
In general, the PR overlap is defined as
\be
Z_{\rm PR}=\langle \Psi| {\cal R}_{\rm PR}|\Psi \rangle,
\label{PRO}
\ee
where $ {\cal R}_{\rm PR}$ is the PR operator that reflects the sites within a segment of lattice
with respect to its central link(s).
Here, we consider the smallest PR segment, i.e., a single cube located at site $\ver=\ver_0$.
Then, ${\cal R}_{\rm PR}$ operates as~\cite{Shapourian1}
\be
&&c_{\ver_0} \rightarrow ic_{\ver_0+\hx+\hy}, \;\; c_{\ver_0+\hx} \rightarrow ic_{\ver_0+\hy},  \nonumber \\
&&c_{\ver_0+\hx+\hy} \rightarrow ic_{\ver_0}, \;\; c_{\ver_0+\hy} \rightarrow ic_{\ver_0+\hx}.
\label{PR}
\ee
Then, the PR overlap of the $1/4$-filling ground state is obtained by calculation the following quantity,
\be
Z^{(1)}_{\rm PR}(1/4)=\langle \Psi_1| {\cal R}_{\rm PR}|\Psi_1 \rangle.
\label{ZPR}
\ee
After some analytical calculation, we obtain
\be
Z^{(1)}_{\rm PR}(1/4)={i \over 2}=e^{i\pi/2}/2.
\label{PRO1}
\ee
Similarly for the state $|\Psi_2\rangle$,
\be
Z^{(2)}_{\rm PR}(1/4)=-{i \over 2}=e^{-i\pi/2}/2.
\label{PRO2}
\ee
The above results of $Z_{\rm PR}(1/4)$ indicates that the $1/4$-filling state 
has $Z_4$-topological phase
corresponding to the phase $\pi/2=(2\pi)/4$, and a single complex fermion emerges per each boundary 
in the OBC by the denominator $2$~\cite{Shapourian1}.
The complex fermion, say at the right boundary, is given by 
$(-\omega^B_{x=L,\hy}-i\tilde{\omega}^A_{x=L,\hy})^\dagger$
coming from $(Q^-_{\ver=(L,0)})^\dagger$.
We think that the emergent $Z_4$-topological phase (not $Z$-topological phase dictated by BDI class)
comes from the four flat-bands structure of the Hamiltonian $H_{\rm SP}$ as shown in Fig.~\ref{Fig5}.
This point will be discussed further in Sec.~VI.

\section{Discussion and Conclusion}

In this paper, by making use of the cube operators, which were heuristically found 
as an extension of the $\ell$-bits (CLS) in Creutz ladder, 
we constructed bilayer flat-band Hamiltonians 
of the exact projective form.
The models have extensive numbers of the LIOMs, thus, full localization occurs in the bulk. 
Since we constructed the Hamiltonians by imposing certain symmetries, time-reversal and chiral symmetries, 
the constructed bilayer flat-band Hamiltonians naturally belong to a symmetric topological class in ten-fold way.
In this work, we explicitly showed that the constructed bilayer flat-band Hamiltonians belong to the BDI class. 
From this classification, the constructed bilayer flat-band Hamiltonians exhibit some topological character, i.e.,
non-trivial bulk topology and presence of the gapless edge modes, in particular in 1D.
The model constructed on a quasi-1D lattice (prism lattice), 
explicitly exhibits SPT phase characterized by topological indexes for periodic system, 
and also the existence of gapless edge modes for the open boundary. 
This is just bulk-edge correspondence in the complete flat-band system.

Here, we would like to give a brief discussion on topological properties of \textit{strongly-localized states}.
For the Hamiltonian $H_{\rm SP}$, the PR overlap shows the existence of the $Z_4$-topology.
It is well known that $Z$-topology of the BDI class by the topological classification actually reduces to 
$Z_8$~\cite{FidKit1,FidKit2}.
For $H_{\rm SP}$, the $Z_4$-topology seems quite plausible as the system has four flat-band structure.
Not only in prism lattice but also in thin cylinder lattice, gapless edge modes are discovered relying on 
the form of the cube operators. 
By the topological classification, 2D systems in the BDI class have no bulk topological properties.
We think that the above results come from the fact that
the localized states in the present models are all described by the $\ell$-bits, and are strictly
confined in a single cube.
Therefore, the spatial dimension does not seem relevant for topological classification in the present systems.
In fact, a close look at the one-dimensional edge modes appearing in the thin cylinder bilayer system
reveals that the edge modes ${\cal B}_{\cal C}$'s are composed of edges modes 
in the the quasi-1D Hamiltonian similar to $H_{\rm SP}$
located in the $y$-direction [see Fig.~\ref{Fig6}].
The chiral symmetry ${\cal S}_xK$ plays an important role for lacing edge modes in the Hamiltonian $H_{\rm SP}$ along the $x$-direction.
This is the mechanism for the emergence of the one-dimensional edge modes, that is, topological
properties of the Hamiltonian $H_{\rm SP}$ and the chiral symmetry collaborate. 
Anyway, careful investigation is required to clarify if this kind of phenomenon is generic or specific.
This is a future work.

\section*{Acknowledgments}
The work is supported in part by JSPS KAKENHI Grant Numbers JP21K13849 (Y.K.). 



\begin{thebibliography}{99}

\bibitem{Smith1}
A. Smith, J. Knolle, D. L. Kovrizhin, and R. Moessner, Phys. Rev. Lett. {\bf 118}, 266601 (2017).

\bibitem{Smith2}
A. Smith, J. Knolle, R. Moessner, and D. L. Kovrizhin, Phys. Rev. Lett. {\bf 119}, 176601 (2017).

\bibitem{WSL2019}
M. Schulz, C.A. Hooley, R. Moessner, F. Pollmann, Phys. Rev. Lett. {\bf 122}, 040606 (2019).

\bibitem{Scherg}
S. Scherg, T. Kohlert, P. Sala, F. Pollmann, H. M. Bharath, I. Bloch, M. Aidelsburger, arXiv:2010.12965 (2020).

\bibitem{McClarty}
P. A. McClarty, M. Haque, A. Sen, and J. Richter, Phys. Rev. B {\bf 102}, 224303 (2020).

\bibitem{Nandkishore}
R. Nandkishore, and D. A. Huse, 
Annual Review of Condensed Matter Physics {\bf 6}, 15 (2015).

\bibitem{Abanin}
D. A. Abanin, E. Altman, I. Bloch, and M. Serbyn, Rev. Mod. Phys. {\bf 91}, 021001 (2019).

\bibitem{Imbrie}
J. Z. Imbrie, V. Ros, and A. Scardicchio, Annalen der Physik {\bf 529}, 1600278 (2017).

\bibitem{Serbyn}
M. Serbyn, Z. Papi\ifmmode \acute{c}\else \'{c}\fi{} , and D. A. Abanin, Phys. Rev. Lett. {\bf 111}, 127201 (2013).

\bibitem{Mukherjee}
S. Mukherjee, M. Di Liberto, P. \"{O}hberg, R. R. Thomson, and N. Goldman, Phys. Rev. Lett. {\bf 121}, 075502 (2018).

\bibitem{Vidal0}
J. Vidal, R. Mosseri, and B.  Dou\c{c}ot, Phys. Rev. Lett. {\bf 81}, 5888 (1998).

\bibitem{Naud}
C. Naud, G. Faini, and D. Mailly, Phys. Rev. Lett. {\bf 86}, 5104 (2001).

\bibitem{Bardarson}
J. H. Bardarson, F. Pollmann, and J. E. Moore, Phys. Rev. Lett. {\bf 109}, 017202 (2012).

\bibitem{Flach}
S. Flach, D. Leykam, J. D. Bodyfelt, P. Matthies, and A. S. Desyatnikov, EPL (Europhysics Letters) {\bf 105}, 30001 (2014).

\bibitem{Mizoguchi2019}
T. Mizoguchi and Y. Hatsugai, EPL (Europhysics Letters) {\bf 127}, 47001 (2019).

\bibitem{KMH2020}
Y. Kuno, T. Mizoguchi, and Y. Hatsugai, Phys. Rev. B {\bf 102}, 241115 (R) (2020).

\bibitem{KMH2020_2}
Y. Kuno, T. Mizoguchi, and Y. Hatsugai, Phys. Rev. A {\bf 102}, 063325 (2020).


\bibitem{Regnault}
N. Regnault and B. A. Bernevig, Phys. Rev. X {\bf 1}, 021014 (2011).

\bibitem{Bergholtz}
E. J. Bergholtz and Z. Liu, Int. J. Mod. Phys. B {\bf 27}, No. 24 1330017, (2013).

\bibitem{Guo}
H. Guo, S. Shen, and S. Feng, Phys. Rev. B {\bf 86}, 085124 (2012).

\bibitem{Budich}
J. C. Budich and E. Ardonne, Phys. Rev. B {\bf 88}, 035139 (2013).

\bibitem{Barbarino_2019}
S. Barbarino, D. Rossini, M. Rizzi, M. Fazio,G. E. Santoro and M. Dalmonte, New J. Phys. {\bf 21}, 043048 (2019).

\bibitem{Tomczak}
P. Tomczak and J. Richter, J. Phys. A: Mathematical and General {\bf 36}, 5399 (2003).

\bibitem{Richter}
J. Richter, J. Schulenburg, P. Tomczak, and D. Schmalfuß, Cond. Matter Phys. {\bf 12}, 507 (2009).

\bibitem{Derzhko}
O. Derzhko and J. Richter, Eur. Phys. J. B {\bf 52}, 23 (2006).

\bibitem{Wildeboer}
J. Wildeboer and A. Seidel, Phys. Rev. B {\bf 83}, 184430 (2011).

\bibitem{Jiang1}
W. Jiang, D. J. P. de Sousa, J. -P. Wang, and T. Low,
Phys. Rev. Lett. {\bf 126}, 106601 (2021).

\bibitem{Jiang2}
W. Jiang, X. Ni, and F. Liu, Accounts of Chemical Research {\bf 54} (2), 416 (2021).

\bibitem{Jiang3}
W. Jiang, Z. Liu, J. -W. Mei, B. Cui and F. Liu, Nanoscale {\bf 11}, 955 (2019).

\bibitem{Creutz1999}
M. Creutz, Phys. Rev. Lett. {\bf 83}, 2636 (1999).

\bibitem{Danieli_1}
C. Danieli, A. Andreanov, and S. Flach, Phys. Rev. B {\bf 102}, 041116 (2020).

\bibitem{Roy}
N. Roy, A. Ramachandran, and A. Sharma, Phys. Rev. Research {\bf 2}, 043395 (2020).

\bibitem{KOI2020}
Y. Kuno, T. Orito, and I. Ichinose, New J. Phys. {\bf 22}, 013032 (2020).

\bibitem{OKI2020}
T. Orito, Y. Kuno, and I. Ichinose, Phys. Rev. B {\bf 101}, 224308 (2020).

\bibitem{OKI2021}
T. Orito, Y. Kuno, and I. Ichinose, Phys. Rev. B {\bf 103}, L060301 (Letter) (2021).

\bibitem{Bermudez}
A. Bermudez, D. Patan\`e, L. Amico, and M. A. Martin-Delgado, Phys. Rev. Lett. {\bf 102}, 135702 (2009).

\bibitem{Junemann}
J. J\"unemann, A. Piga, L. Amico, S.-J. Ran, M. Lewenstein, M. Rizzi, and A. Bermudez, Phys. Rev. X {\bf 7}, 031057 (2017).

\bibitem{Sun}
N. Sun, and L.-K. Lim, Phys. Rev. B {\bf 96}, 035139 (2017).

\bibitem{Zurita}
J. Zurita, C. E. Creffield, and G. Platero, Advanced Quantum Technologies {\bf 3}, 1900105 (2020).

\bibitem{YK2020}
Y. Kuno, Phys. Rev. B {\bf 101}, 184112 (2020).

\bibitem{Chiu}
C. K. Chiu, J. C. Y. Teo, A. P. Schnyder, and S. Ryu, Rev. Mod. Phys. {\bf 88}, 035005 (2016).

\bibitem{Ludwig}
A. W. W. Ludwig, Phys. Scr. {\bf T168}, 014001 (2016).

\bibitem{Altland}
A. Altland and M. R. Zirnbauer, Phys. Rev. B {\bf 55}, 1142 (1997).

\bibitem{Ryu}
S. Ryu, A. P. Schyder, A. Furusaki, and A. W. W. Ludwig, New J. Phys. {\bf 12}, 065010 (2010).

\bibitem{Pollmann2010}
F. Pollmann, A. M. Turner, E. Berg, and M. Oshikawa, 
Phys. Rev. B {\bf 81}, 064439 (2010).

\bibitem{Chen}
X. Chen, Z.-C. Gu, and X.-G. Wen, Phys. Rev. B {\bf 84}, 235128 (2011).

\bibitem{Pollmann2012}
F. Pollmann, E. Berg, A. M. Turner, and M. Oshikawa, Phys. Rev. B {\bf 85}, 075125 (2012).

\bibitem{Bahri}
Y. Bahri, R. Vosk, E. Altman, and A. Vishwanath, Nat. Commun. {\bf 6}, 7341 (2015).

\bibitem{SachdevM}
S. Sachdev and J. Ye, Phys. Rev. Lett. {\bf 70}, 3339 (1993).

\bibitem{YeM}
S. Sachdev, Phys. Rev. X {\bf 5}, 041025 (2015).

\bibitem{KitaevM}
A. Kitaev, talk at KITP Program: Entanglement in Strongly-Correlated Quantum Matter (2015).

\bibitem{Asboth}
J. K. Asboth, L. Oroszlany, and A. Palyi, \textit{A Short Course on Topological Insulators: Band-structure Topology and Edge States in One and Two Dimensions} (Springer, Berlin, 2016).

\bibitem{Zak1}
J. Zak, Phys. Rev. {\bf 134}, A1602 (1964). 
\bibitem{Zak2}
J. Zak, Phys. Rev. {\bf 134}, A1607 (1964).

\bibitem{Fang}
C. Fang, M. J. Gilbert, and B. A. Bernevig, Phys. Rev. B {\bf 86}, 115112 (2012).

\bibitem{You2017}
Y.-Z. You, A. W. W. Ludwig, and C. Xu, Phys. Rev. B {\bf 95}, 115150 (2017). 

\bibitem{Quspin}
We employed the Quspin solver: P. Weinberg and M. Bukov, SciPost Phys. {\bf 7}, 20 (2019); {\bf 2}, 003 (2017).

\bibitem{Hatsugai2005}
Y. Hatsugai, J. Phys. Soc. Jpn. {\bf 74}, 1374 (2005).

\bibitem{Hatsugai2006}
Y. Hatsugai, J. Phys. Soc. Jpn. {\bf 75}, 123601 (2006).

\bibitem{Hatsugai2007}
Y. Hatsugai, J. Phys. Condens. Matter {\bf 19}, 145209 (2007).

\bibitem{Hirano2008}
T. Hirano, H. Katsura, and Y. Hatsugai, Phys. Rev. B {\bf 77}, 094431 (2008). 

\bibitem{Katsura2007}
H. Katsura, T. Hirano, and Y. Hatsugai, Phys. Rev. B {\bf 76}, 012401 (2007).

\bibitem{HM2011}
Y. Hatsugai and I. Maruyama, EPL (Europhysics Letters) {\bf 95}, 20003 (2011).

\bibitem{Kitaev}
A. Y. Kitaev, Physics-Uspekhi {\bf 44}, 131 (2001).

\bibitem{Fendley}
P. Fendley, Journal of Statistical Mechanics: Theory and Experiment {\bf 2012} P11020, (2012).
\bibitem{McGinley}
M. McGinley, J. Knolle, and A. Nunnenkamp, Phys. Rev. B {\bf 96}, 241113 (2017).


\bibitem{Guo}
H. Guo, Phys. Rev. A {\bf 86}, 055604 (2012).

\bibitem{PollTur}
F. Pollmann and A. M. Tuener, Phys. Rev. B {\bf 86}, 125441 (2012).

\bibitem{Shapourian1}
H. Shapourian, K. Shiozaki, and S. Ryu, Phys. Rev. Lett. {\bf 118}, 216402 (2017).

\bibitem{FidKit1}
L. Fidkowski and A. Kitaev, Phys. Rev. B {\bf 81}, 134509 (2010).

\bibitem{FidKit2}
L. Fidkowski and A. Kitaev, Phys. Rev. B {\bf 83}, 075103 (2011).



\end{thebibliography}
\end{document}